\documentclass[showpacs,twocolumn,prd,aps]{revtex4}
\usepackage{amsfonts}
\usepackage{amsmath}
\usepackage{amsbsy}
\usepackage{amssymb}

\font\grb=eurb10

\def\bchi{\hbox{\grb\char'037}\,}

\begin{document}

\title{Integrability of the symmetry reduced bosonic dynamics\\ and soliton generating transformations in the low\\ energy heterotic string effective theory}
\author{G.A.~Alekseev}
     \email{G.A.Alekseev@mi.ras.ru}
\affiliation{\centerline{
\hbox{Steklov Mathematical
Institute of the Russian Academy of Sciences,}}\\
\centerline{\hbox{Gubkina str. 8, 119991, Moscow, Russia}}
}

\begin{abstract}
\noindent
Integrable structure of the symmetry reduced dynamics of massless bosonic sector of the heterotic string effective action is presented. For string background equations that govern in the space-time of $D$ dimensions ($D\ge 4$) the dynamics of interacting gravitational, dilaton,
antisymmetric tensor and any number $n\ge 0$ of Abelian vector gauge fields, all depending only on two coordinates, we construct an \emph{equivalent} $(2 d+n)\times(2 d+n)$ matrix spectral problem ($d=D-2$). This spectral problem provides the base for the development of various solution constructing procedures (dressing transformations, integral equation methods). For the case of the absence of Abelian
gauge fields, we present the soliton generating transformations of any background with interacting gravitational, dilaton and the second rank antisymmetric tensor fields. This new soliton generating procedure is available for constructing of various types of field
configurations including stationary axisymmetric fields, interacting plane, cylindrical or some other types of waves and cosmological solutions.

\end{abstract}


\maketitle

\section*{Introduction}
In recent decades, development of the (super)string theory toward a consistent description
of all interactions suggested  fundamental changes of our view of the picture of space-time
and field dynamics there \cite{Green-Schwarz-Witten:1987, Horowitz:2005}. Many interesting
features of this picture  have been discovered  using the solutions of the  corresponding
low-energy effective theories. These solutions play an important role in the analysis of
various nonperturbative aspects of the string theory (see \cite{Youm:1999, Mohaupt:2000,
Emparan-Reall:2008} and the references therein). However, most of these solutions were
found using some very particular ans\"atze, or global symmetry transformations. More
systematic approaches and flexible methods may arise for symmetry reduced field
equations, provided these occur to be integrable. This integrability may lead to constructing
of large varieties of multiparametric solutions for physically different types of interacting
fields.

The symmetry reduced equations of pure vacuum Einstein gravity in $D$ dimensions are clearly integrable and for their solution one can use (without substantial changes) the inverse scattering approach developed for $D=4$ space-times thirty years ago by Belinski and Zakharov \cite{Belinski-Zakharov:1978}. Many authors used this approach to construct soliton solutions in four and five dimensions  (e.g., \cite{Belinski-Ruffini:1980,Pomeransky:2006,
Elvang-Figueras:2007, Elvang-Rodriguez:2007}).

In $D=4$ space-times with two commuting isometries,
the dynamics of dilaton and axion fields coupled to gravity
also was described using the Belinski and Zakharov inverse scattering approach \cite{Bakas:1996}. In higher dimensions or/and in the presence of vector gauge fields, this dynamics have been studied by many authors, who asserted its integrability, however, understanding of the former has not been  enough to give rise to some solution generating methods beyond global symmetry transformations \cite{Kechkin:2004}.

In this paper, a completely  integrable structure of the dynamics of massless bosonic sector of  heterotic string effective action for space-times with $D\ge 4$ dimensions and with $d=D-2$  commuting isometries  is described. For interacting gravitational, dilaton, second rank antisymmetric tensor and any number $n\ge 0$ of Abelian gauge vector fields we construct an \emph{equivalent} $(2 d+n)\times(2 d+n)$-matrix spectral problem which
provides the base for development of various solution constructing procedures.

For vanishing gauge vector fields, the soliton generating transformations are constructed here using an appropriate modification of the construction \cite{Alekseev:1980} of Einstein - Maxwell solitons. For vacuum fields, the relations between our spectral problem  and  Belinski-Zakharov one and corresponding vacuum solitons is shown explicitly.

\section*{Massless bosonic modes of heterotic string theory}
The massless bosonic part of heterotic string effective action in space-time with $D\ge 4$ in the string frame is
\begin{equation}
\label{String_Frame}
\begin{array}{l}
{\cal S}=\!\!\displaystyle\int e^{-\widehat{\Phi}}\left\{\widehat{R}{}^{(D)}+\nabla_M \widehat{\Phi} \nabla^M\widehat{\Phi}
-\displaystyle\frac 1{12} H_{MNP} H^{MNP}\right.\\[1ex]
\phantom{{\cal S}^{(D)}=\int e^{-\widehat{\widehat{\Phi}}}}\left.
-\dfrac 12\sum\limits_{\mathfrak{p}=1}^n\displaystyle F_{MN}{}^{(\mathfrak{p})} F^{MN\,(\mathfrak{p})}\right\}\sqrt{- \widehat{G}}\,d^{D}x
\end{array}
\end{equation}
where $M,N,\ldots=1,2,\ldots,D$ and $\mathfrak{p}=1,\ldots n$; $\widehat{G}{}_{MN}$ possesses the ``most positive'' Lorentz signature. Metric $\widehat{G}{}_{MN}$ and dilaton field $\widehat{\Phi}$ are related to the metric  $G_{MN}$ and dilaton  $\Phi$ in the Einstein frame as
\begin{equation}
\label{Einstein_Frame}
\widehat{G}{}_{MN}=e^{2\Phi} G_{MN},\qquad
\widehat{\Phi}=(D-2)\Phi.
\end{equation}
The components of a three-form $H$ and two-forms $F{}^{(\mathfrak{p})}$ are determined in terms of antisymmetric tensor field $B_{MN}$ and Abelian gauge field potentials $A_M{}^{(\mathfrak{p})}$ as
\[\begin{array}{l}
H_{MNP}=3\bigl(\partial_{[M} B{}_{NP]}-\sum\limits_{\mathfrak{p}=1}^n A{}_{[M}{}^{(\mathfrak{p})} F_{NP]}{}^{(\mathfrak{p})}\bigr),\\[0.5ex]
F_{MN}{}^{(\mathfrak{p})}=2\,\partial_{[M} A{}_{N]}{}^{(\mathfrak{p})},\qquad B_{MN}=-B_{NM}.
\end{array}
\]

\section*{Space-time symmetry ansatz}
We consider the space-times with $D\ge 4$ dimensions which admit $d=D-2$ commuting
Killing vector fields. All field components and potentials are assumed to be functions of
only two coordinates  $x^1$ and $x^2$, one of which can be timelike or both are spacelike
coordinates. We assume also the following structure of metric components
\begin{equation}\label{Metric}
G_{MN}=\begin{pmatrix}g_{\mu\nu}&0\\
0 & G_{ab}
\end{pmatrix}\qquad
\begin{array}{l}
\mu,\nu,\ldots=1,2\\
a,b,\ldots=3,4,\ldots D
\end{array}
\end{equation}
while the components of field potentials take the forms
\begin{equation}\label{BAfields}
B_{MN}=\begin{pmatrix} 0& 0\\
0 & B_{ab}
\end{pmatrix},\qquad
A_M{}^{(\mathfrak{p})}=\begin{pmatrix}0\\ A_a{}^{(\mathfrak{p})}
\end{pmatrix}.
\end{equation}
We chose $x^1$, $x^2$ so that $g_{\mu\nu}$ takes a conformally flat form
\[g_{\mu\nu}=f \eta_{\mu\nu},\qquad\eta_{\mu\nu}=
\left(\begin{array}{ll}\epsilon_1 & 0\\
0&\epsilon_2\end{array}\right),\qquad\begin{array}{l}\epsilon_1=\pm1\\
\epsilon_2=\pm1\end{array}\]
where $f(x^\mu)>0$ and  the sign symbols $\epsilon_1$ and $\epsilon_2$ allow to consider various types of fields. The field equations imply that the function $\alpha(x^1,x^2)>0$ is  ``harmonic'' one:
\[
\det\Vert G_{ab}\Vert\equiv \epsilon\alpha^2, \qquad \eta^{\mu\nu}\partial_\mu\partial_\nu\alpha=0,\qquad\epsilon=-\epsilon_1 \epsilon_2.
\]
where $\eta^{\mu\nu}$ is inverse to $\eta_{\mu\nu}$, and
therefore, the function $\beta(x^\mu)$ can be defined as ``harmonically'' conjugated  to $\alpha$:
\[\partial_\mu\beta=\epsilon\varepsilon_\mu{}^\nu\partial_\nu\alpha,
\qquad
\varepsilon_\mu{}^\nu=\eta_{\mu\gamma}\varepsilon^{\gamma\nu},\qquad
\varepsilon^{\mu\nu}=\begin{pmatrix}0&1\\-1&0\end{pmatrix}.
\]
Using the functions $(\alpha,\beta)$, we construct a pair $(\xi,\eta)$ of real null coordinates in the hyperbolic case or complex conjugated to each other coordinates in the elliptic case:
\[\left\{\begin{array}{l} \xi=\beta+j\alpha,\\[1ex]
\eta=\beta-j\alpha,\end{array}\right.\quad
j=\left\{\begin{array}{llll}
1,&\epsilon=1&-&\hbox{hyperbolic case},\\[1ex]
i,&\epsilon=-1&-&\hbox{elliptic case.}
\end{array}\right.
\]
In particular, for stationary axisymmetric fields $\xi=z+i\rho$,
$\eta=z-i\rho$, whereas for plane waves or for cosmological solutions $\xi=-x+t$, $\eta=-x-t$ or these may have more complicate expressions in terms of $x^1$, $x^2$.

\section*{Dynamical equations}
The symmetry reduced dynamical equations can be presented in the form of a matrix analogue of the known Ernst equations \footnote{For stationary fields, in different notations the matrix Ernst equations were found earlier (see the survey \cite{Kechkin:2004}).} expressed in terms of the string frame variables -- a symmetric $d\times d$-matrix $\mathcal{G}=e^{2\Phi}\Vert G_{ab}\Vert$, antisymmetric $d\times d$-matrix $\mathcal{B}=\Vert B_{ab}\Vert$, a rectangular $d\times n$-matrix $\mathcal{A}=\Vert A_a{}^{(\mathfrak{p})}\Vert$ and a scalar $\widehat{\Phi}$:
\begin{equation}\label{Ernst_equations}
\left\{\begin{array}{l}
\eta^{\mu\nu}\partial_\mu
(\alpha \partial_\nu{\cal E})- \alpha\,\eta^{\mu\nu}(\partial_\mu{\cal E}-2\partial_\mu {\cal A}{\cal A}^T)\mathcal{G}^{-1}\partial_\nu{\cal E}=0,
\\[1ex]
\eta^{\mu\nu}\partial_\mu(\alpha \partial_\nu{\cal A})-\alpha\,\eta^{\mu\nu}
(\partial_\mu{\cal E}-2\partial_\mu {\cal A}{\cal A}^T)\mathcal{G}^{-1}\partial_\nu{\cal A}=0,\\[1ex]
\eta^{\mu\nu}\partial_\mu\partial_\nu\alpha=0,
\end{array}\right.
\end{equation}
where $T$ means a matrix transposition and
\begin{equation}\label{Ernst_potential}
{\cal E}=\mathcal{G}+\mathcal{B}+{\cal A} {\cal A}^T,
\qquad
\det \mathcal{G}=\epsilon\alpha^2e^{2\widehat{\Phi}}.
\end{equation}
Equations (\ref{Ernst_equations}) imply the existence of two other matrix potentials.
These are the antisymmetric $d\times d$-matrix potential $\widetilde{\mathcal{B}}$ and
$d\times n$-matrix potential $\widetilde{\mathcal{A}}$ defined as
\begin{equation}\label{duals}
\begin{array}{l}
\partial_\mu\widetilde{\mathcal{B}}= -\epsilon\alpha\varepsilon_\mu{}^\nu \mathcal{G}^{-1}(\partial_\nu\mathcal{B}-\partial_\nu{\cal A} {\cal A}^T+{\cal A} \partial_\nu{\cal A}^T)\mathcal{G}^{-1},\\[1ex]
\partial_\mu\widetilde{\mathcal{A}}=-\epsilon\alpha \varepsilon_\mu{}^\nu \mathcal{G}^{-1}\partial_\nu \mathcal{A}+\widetilde{\mathcal{B}}\,\partial_\mu \mathcal{A}.
\end{array}
\end{equation}
The remaining part of the field equations does not have a dynamical character and it allows to calculate the conformal factor $f$ in quadratures (see the expressions below), provided the solution of dynamical equations was found.

\section*{``Null curvature'' representation}
It is remarkable that the dynamical equations (\ref{Ernst_equations}) can be transformed into a pair of first order matrix equations for two $(2 d+n)\times (2 d+n)$-matrix functions $\mathbf{U}$ and $\mathbf{V}$ which are real (in the hyperbolic case) or complex conjugated to each other (in the elliptic case):
\begin{equation}\label{UVequations}
\partial_\eta \mathbf{U}+\partial_\xi \mathbf{V}+\dfrac {[\mathbf{U},\mathbf{V}]}{\xi-\eta}
=0,\quad \partial_\eta \mathbf{U}-\partial_\xi \mathbf{V}=0,
\end{equation}
where $\mathbf{U}$ and $\mathbf{V}$ possess the $3\times 3$ block-matrix structures
\[\begin{array}{l}
\mathbf{U}=\begin{pmatrix}
I_d&0&0\\[1ex]
B_+& I_d&0\\[1ex]
C_+&0&I_n
\end{pmatrix}
\begin{pmatrix}
I_d&-{\cal E}_\xi&-2{\cal A}_\xi\\[1ex]
0& 0&0\\[1ex]
0&0&0
\end{pmatrix}
\begin{pmatrix}
I_d&0&0\\[1ex]
-B_+& I_d&0\\[1ex]
-C_+&0&I_n
\end{pmatrix}\\[5ex]
\mathbf{V}=\begin{pmatrix}
I_d&0&0\\[1ex]
B_-& I_d&0\\[1ex]
C_-&0&I_n
\end{pmatrix}
\begin{pmatrix}
I_d&-{\cal E}_\eta&-2{\cal A}_\eta\\[1ex]
0& 0&0\\[1ex]
0&0&0
\end{pmatrix}
\begin{pmatrix}
I_d&0&0\\[1ex]
-B_-& I_d&0\\[1ex]
-C_-&0&I_n
\end{pmatrix}
\end{array}
\]
in which the subscripts $\xi$ and $\eta$ denote partial derivatives, $I_d$ and $I_n$ mean $d\times d$ and $n\times n$ unit matrices respectively. The expressions for $d\times d$-blocks $B_\pm$ and $n\times d$-blocks $C_\pm$ include the potentials defined in (\ref{duals}):
\begin{equation}\label{BC_matrices}\begin{array}{lcccl}
B_+=\widetilde{\mathcal{B}}-j\alpha \mathcal{G}^{-1}&&&&
C_+=-(\widetilde{\mathcal{A}}{}^T+\mathcal{A}{}^T B_+)
\\[1ex]
B_-=\widetilde{\mathcal{B}}+j\alpha \mathcal{G}^{-1}&&&&
C_-=-(\widetilde{\mathcal{A}}{}^T+\mathcal{A}{}^T B_-)
\end{array}
\end{equation}
In these coordinates and notations, the second relation in (\ref{duals}) is equivalent to $\partial_\xi \widetilde{\mathcal{A}}=B_+\partial_\xi\mathcal{A}$,\quad $\partial_\eta \widetilde{\mathcal{A}}=B_-\partial_\eta\mathcal{A}$.

\section*{Spectral problem}
To obtain a spectral problem equivalent to the dynamical equations, we construct a linear system with a free complex (``spectral") parameter $w\in \overline{\mathbb{C}}$ and with the integrability conditions (\ref{UVequations}) and supply it with the constraints providing $\mathbf{U}$ and $\mathbf{V}$ to have the above structures and satisfy  (\ref{Ernst_potential}) -- (\ref{BC_matrices}). In our spectral problem it is required to find four $(2 d+n)\times (2 d+n)$-matrix functions
\begin{equation}\label{PsiUVW}
\mathbf{\Psi}(\xi,\eta,w),\,\,\mathbf{U}(\xi,\eta),\,\,
\mathbf{V}(\xi,\eta),\,\,\mathbf{W}(\xi,\eta,w)
\end{equation}
which should satisfy the following linear system with algebraic constraints on its matrix coefficients
\begin{equation}\label{LinSys}
\left\{\begin{array}{l}
2(w-\xi)\partial_\xi \mathbf{\Psi}=\mathbf{U}(\xi,\eta) \mathbf{\Psi}\\[2ex]
2(w-\eta)\partial_\eta \mathbf{\Psi}=\mathbf{V}(\xi,\eta) \mathbf{\Psi}
\end{array}\hskip1ex\right\Vert
\hskip1ex\begin{array}{l}
{\bf U}\cdot{\bf U} ={\bf U},\hskip1ex \text{tr}{\bf U}=d \\[2ex]
{\bf V}\cdot{\bf V} ={\bf V},\hskip1ex\text{tr}{\bf V}=d
\end{array}
\end{equation}
Besides that, it is required that the system (\ref{LinSys}) possesses a symmetric matrix integral $\mathbf{K}(w)$ such that
\begin{equation}\label{W_Condition}
\left\{\begin{array}{l}
{\bf \Psi}^T {\bf W} {\bf \Psi}={\bf K}(w)\\[2ex]
{\bf K}^T(w)={\bf K}(w)
\end{array}
\quad\right\Vert\quad
\dfrac{\partial\mathbf{W}}{\partial w}=\mathbf{\Omega},\quad
\mathbf{\Omega}=\begin{pmatrix}
0&I_d&0\\I_d&0& 0\\0&0&0
\end{pmatrix}
\end{equation}
where $\mathbf{\Omega}$ is $(2 d+n)\times (2 d+n)$ matrix. We require also
\begin{equation}\label{reality}
\overline{\mathbf{\Psi}(\xi,\eta,w)}= \mathbf{\Psi}(\xi,\eta,\overline{w}),\hskip1ex
\overline{\mathbf{K}(w)}= \mathbf{K}(\overline{w}),\hskip1ex
\mathbf{W}_{(3)(3)}=I_n.
\end{equation}
where $\mathbf{W}_{(3)(3)}$  is the lower right $n\times n$ block of $\mathbf{W}$.
In accordance with  (\ref{PsiUVW})--(\ref{W_Condition}), $\mathbf{W}_{(3)(3)}$ is a constant matrix and therefore, the condition $\mathbf{W}_{(3)(3)}=I_n$ can be always achieved by an appropriate gauge transformation.

The equivalence of the spectral problem (\ref{PsiUVW})--(\ref{reality}) to (\ref{Ernst_equations}) can be seen from a direct calculation similar to \cite{Alekseev:2005b}.

\subsection*{Field variables and potentials}
The conditions (\ref{PsiUVW})--(\ref{reality}) imply, that   $\mathbf{W}$ has the form
\begin{equation}\label{GW}
\begin{array}{l}
\mathbf{W}=(w-\beta)\mathbf{\Omega}+\mathbf{G},\\[1ex]
\mathbf{G}=\begin{pmatrix}
\epsilon\alpha^2\mathcal{G}^{-1}-\widetilde{\mathcal{B}} \mathcal{G}\widetilde{\mathcal{B}} +\widetilde{\mathcal{A}}\widetilde{\mathcal{A}}{}^T
&\widetilde{\mathcal{B}}\mathcal{G}+ \widetilde{\mathcal{A}}\mathcal{A}{}^T
&\widetilde{\mathcal{A}}\\[1ex]
-\mathcal{G} \widetilde{\mathcal{B}}
+{\mathcal{A}}\widetilde{\mathcal{A}}{}^T
&\mathcal{G}+\mathcal{A}\mathcal{A}{}^T& \mathcal{A}\\[1ex]
\widetilde{\mathcal{A}}{}^T&\mathcal{A}^T&I_n
\end{pmatrix}
\end{array}
\end{equation}
where $(2 d+n)\times (2 d+n)$ matrix $\mathbf{G}$ is real and symmetric, that
its $d\times d$ matrix blocks $\mathcal{G}$, $\widetilde{\mathcal{B}}$
are symmetric and antisymmetric respectively and that these matrix variables,
together with $d\times n$ matrices $\mathcal{A}$ and $\widetilde{\mathcal{A}}$, satisfy
(\ref{Ernst_equations}) -- (\ref{duals}).

The conformal factor $\widehat{f}=e^{2\Phi} f$ in the Weyl form of conformally flat part of string frame metric $\widehat{f}(d\alpha^2-\epsilon\, d\beta^2)$ can be calculated in quadratures (``$\text{tr\,}$" denotes a trace):
\[\left\{\begin{array}{l}
\partial_\xi\log\left(
\alpha^{d/2} e^{-\widehat{\Phi}} \widehat{f}\right)= -\dfrac 18
\text{tr}\left[(\mathbf{G}^{-1}+\mathbf{E}) \mathbf{U}^T\mathbf{\Omega}\mathbf{U}\right]\\[1ex]
\partial_\eta\log\left(\alpha^{d/2} e^{-\widehat{\Phi}} \widehat{f}\right)= -\dfrac 18
\text{tr}\left[(\mathbf{G}^{-1}+\mathbf{E}) \mathbf{V}^T\mathbf{\Omega}\mathbf{V}\right]
\end{array}\right.
\]
where $(2 d+n)\times(2 d+n)$ matrix $\mathbf{E}=\mathbf{I}-\mathbf{\Omega}^2$.

\subsection*{Global symmetries}
The spectral problem constructed above admits global symmetry transformations (including the discrete ones)
\begin{equation}\label{Symmetries}
\left.\begin{array}{l}
\mathbf{U}\to\mathbf{A} \mathbf{U}\mathbf{A}^{-1},\quad
\mathbf{V}\to\mathbf{A}\mathbf{V}\mathbf{A}^{-1}\\[1ex]
\mathbf{\Psi}\to \mathbf{A}\mathbf{\Psi},\quad
\mathbf{W}\to(\mathbf{A}^T){}^{-1}\mathbf{W}\mathbf{A}^{-1}
\end{array}\hskip1ex\right\Vert\hskip1ex
\begin{array}{l}
\mathbf{A}^T\mathbf{\Omega}\mathbf{A}=\mathbf{\Omega}\\[1ex]
\mathbf{W}_{(3)(3)}\equiv I_n
\end{array}
\end{equation}
where the real constant matrix $\mathbf{A}$ is determined by two invariance conditions shown just above on the right.  Some of these symmetries are not pure gauge and generate physically different solutions from a given one.

\subsection*{Soliton generating transformations with $\mathcal{A}\equiv 0$}
For vanishing vector gauge fields ($\mathcal{A}\equiv 0$), the problem (\ref{PsiUVW})--(\ref{reality}) reduces to $2 d\times 2 d$ matrix form which admits the soliton generating transformations. Given a solution of (\ref{Ernst_equations}) with $\mathcal{A}\equiv 0$, we  denote its $2 d\times 2 d$ matrices by "$\circ$". For
the one-soliton solution on this background we assume
\[
\mathbf{\Psi}=\bchi\, {\overset \circ \Psi},\quad \bchi=\mathbf{I}+\dfrac {\mathbf{R}(\xi,\eta)}{w-w_1}, \quad \bchi^{-1}=\mathbf{I}+\dfrac {\mathbf{S}(\xi,\eta)}{w-\widetilde{w}_1}
\]
where $w_1$, $\widetilde{w}_1$ are real constants and  $2 d\times 2 d$-matrices $\mathbf{R}$, $\mathbf{S}$ depend on $\xi$, $\eta$ only. For consistency we also assume
\[\mathbf{K}(w)=\left(\dfrac{w-\widetilde{w}_1}{w-w_1}\right) {\overset \circ {\mathbf{K}}}(w).
\]
Then the conditions (\ref{PsiUVW})--(\ref{reality}) imply that for this soliton solution
$\mathbf{U}$, $\mathbf{V}$ and $\mathbf{W}$ take the forms
\[
\begin{array}{l}
\mathbf{W}={\overset \circ {\mathbf{W}}}-\mathbf{\Omega}\cdot \mathbf{R}-\mathbf{R}^T\cdot\mathbf{\Omega} +(w_1-\widetilde{w}_1) \mathbf{\Omega},\\
\mathbf{U}={\overset \circ {\mathbf{U}}}+2\partial_\xi \mathbf{R},\quad \mathbf{V}={\overset \circ {\mathbf{V}}}+2\partial_\eta \mathbf{R},\\
[1ex]
\mathbf{R}=(w_1-\widetilde{w}_1)\, \mathbf{p}\cdot(\mathbf{m}\cdot\mathbf{p})^{-1}
\cdot \mathbf{m}
\end{array}
\]
in which $(d\times 2 d)$ matrix $\mathbf{m}$ and $(2 d\times d)$ matrix $\mathbf{p}$ are
determined in terms of the background solution as
\begin{equation}\label{mpmatrices}
\left.\begin{array}{l}
\mathbf{m}=\mathbf{k}\cdot{\overset \circ {\mathbf{\Psi}}}{}^{-1}(\xi,\eta,w_1)\\
\mathbf{p}={\overset \circ {\mathbf{\Psi}}}(\xi,\eta,\widetilde{w}_1)\cdot
\mathbf{l}
\end{array}\hskip1ex\right\Vert\hskip1ex
\begin{array}{l}
\mathbf{k}\cdot {\overset \circ {\mathbf{K}}}{}^{-1}(w_1)\cdot \mathbf{k}^T=0,\\
\mathbf{l}^T\cdot {\overset \circ {\mathbf{K}}}(\widetilde{w}_1)\cdot \mathbf{l}=0.
\end{array}
\end{equation}
with the "integration constants " -- the real $(d\times 2 d)$ matrix $\mathbf{k}$ and
$(2 d\times d)$ matrix $\mathbf{l}$, which must have the ranks equal $d$ and satisfy the
algebraic constraints (\ref{mpmatrices}). To solve  (\ref{mpmatrices}), we note that
$\mathbf{R}$ and $\mathbf{W}$ remain unchanged if we multiply  $\mathbf{k}$ from the
left and $\mathbf{l}$ from the right by some constant nondegenerate $d\times d$ matrices.
Therefore, without loss of generality, we can put some $d$ columns of $\mathbf{k}$ and
$d$ rows of $\mathbf{l}$ equal to $d\times d$ unit matrices. Then, for example, a
transformation ${\overset \circ {\mathbf{\Psi}}}\to {\overset \circ {\mathbf{\Psi}}}\cdot
\mathbf{C}(w)$, such that ${\overset \circ {\mathbf{K}}}\to \mathbf{C}^T\cdot{\overset \circ
{\mathbf{K}}}\cdot \mathbf{C}=\mathbf{\Omega}$, linearizes the  constraints
(\ref{mpmatrices}). This procedure can be generalized to any number of solitons.

\subsection*{On the Belinski-Zakharov vacuum solitons}
For vacuum fields in $D$ dimensions, the fundamental solution $\mathbf{\Psi}(\xi,\eta,w)$ of our spectral problem takes the form
\[\mathbf{\Psi}=\dfrac 1{\lambda^2-\epsilon \alpha^2} \begin{pmatrix} \lambda\,\,I_d&-\mathcal{G}\\[1ex]
-\epsilon \alpha^2 \mathcal{G}^{-1}&\lambda\,\,I_d \end{pmatrix}
\begin{pmatrix}
\psi_{\scriptscriptstyle{BZ}}&0\\[1ex] 0& \lambda (\psi_{\scriptscriptstyle{BZ}}^{-1})^T
\end{pmatrix}
\]
where $2w=\lambda+2\beta+\epsilon\alpha^2/\lambda$ and
$\psi_{\scriptscriptstyle{BZ}}(\xi,\eta,\lambda)$ is a fundamental solution of Belinski-Zakharov vacuum $d\times d$-matrix spectral problem. The poles
$\lambda=\mu_k(\xi,\eta)$ of $\psi_{\scriptscriptstyle{BZ}}$ on the $\lambda$ plane correspond
to poles $w=w_k$ of $\mathbf{\Psi}(\xi,\eta,w)$ on the
$w$ plane ($2w_k=\mu_k+2\beta+\epsilon\alpha^2/\mu_k$) and therefore, the known vacuum solitons are also soliton solutions for our spectral problem (sometimes, however, of a more general type). For any $D \ge 4$ and for different backgrounds, our soliton generating procedure leads to large families of solutions for interacting fields whose physical and geometrical properties need further investigation.
\bigskip
\bigskip
\bigskip

\subsection*{Monodromy transform approach for $\mathcal{A}\ne 0$}
\vspace{-1ex}
The spectral problem (\ref{PsiUVW})--(\ref{reality}) possesses an important monodromy preserving property providing the base for application of the so called monodromy transform approach suggested in \cite{Alekseev:1985, Alekseev:1988} (see also
\cite{Alekseev:2001}) for Einstein-Maxwell fields. It allows to reduce our spectral problem to the equivalent system of linear singular integral equations, which admits an explicit calculation of infinite hierarchies of solutions for any (analytically matched) rational monodromy data. However, such developments are expected to be the subject of
subsequent publications.

\subsection*{Acknowledgements}

This work was supported in parts by the Russian Foundation for Basic Research (grants 08-01-00501, 08-01-00618, and 09-01-92433-CE) and the programs "Mathematical Methods of Nonlinear Dynamics" of  Russian Academy of Sciences
and "Leading Scientific Schools" of Russian Federation, (grant NSh-1959.2008.1).

\end{document}